\documentclass[reprint,amssymb,pra,longbibliography]{revtex4-2}
\usepackage[utf8]{inputenc}
\usepackage[T1]{fontenc}
\usepackage{graphicx}
\usepackage{amsmath}
\usepackage{amssymb}
\usepackage{mathrsfs}
\usepackage{braket}
\usepackage{geometry}
\usepackage{here}
\usepackage{lmodern}
\usepackage{array}
\usepackage{natbib}
\usepackage{url}
\usepackage{textcase}
\usepackage{bm}
\usepackage{natbib}
\usepackage[usenames,dvipsnames]{color}
\usepackage{hyperref}
\hypersetup{
colorlinks=true,
linkcolor=blue,
citecolor=blue,
filecolor=green,
urlcolor=blue,
}

\begin{document}


\title{Quantum spin state selectivity and magnetic tuning of ultracold chemical reactions of triplet alkali-metal dimers with alkali-metal atoms}

\author{Rebekah Hermsmeier$^{1}$, {Jacek K{\l}os$^{2,3}$, Svetlana Kotochigova$^{3}$ and Timur V. Tscherbul$^{1}$} }

\affiliation{$^{1}$Department of Physics, University of Nevada, Reno, NV, 89557, USA\\
$^{2}$Department of Physics, Joint Quantum Institute, University of Maryland College Park, College Park, Maryland, 20742, USA\\
$^{3}$Department of Physics, Temple University,  Philadelphia, PA 19122, USA}

\date{\today}

\begin{abstract}


We demonstrate that it is possible to efficiently control ultracold chemical reactions of alkali-metal atoms
colliding with open-shell alkali-metal dimers in their metastable triplet states by choosing the internal hyperfine and
rovibrational states of the reactants as well as by inducing magnetic Feshbach resonances with an external
magnetic field.
We base these conclusions on coupled-channel statistical calculations that include the effects of hyperfine
contact and magnetic-field-induced Zeeman interactions on ultracold chemical reactions of hyperfine-resolved
ground-state Na and the triplet NaLi(a$^3\Sigma^+$) producing singlet Na$_2$($^1\Sigma^+_g$) and a Li
atom. We find that the reaction rates are sensitive to the initial hyperfine states of the reactants. The chemical reaction
of fully spin-polarized, high-spin states of rotationless  NaLi(a$^3\Sigma^+, v = 0, N = 0$) molecules with fully
spin-polarized Na is suppressed by a factor of 10-100 compared to that of unpolarized reactants.
We interpret these findings within the adiabatic state model, which treats the reaction as a sequence of
nonadiabatic transitions between the initial non-reactive high-spin state and the final low-spin states of the
reaction complex. In addition, we show that magnetic Feshbach resonances can similarly change reaction rate
coefficients by several orders of magnitude. Some of these resonances are due to resonant trimer bound states
dissociating to the  $N=2$ rotational state of NaLi(a$^3\Sigma^+, v = 0$) and would thus exist in systems
without hyperfine interactions.


\end{abstract}

\maketitle

{\it Introduction.}
Recent experimental advances in molecular cooling and trapping have opened up new avenues  of research into  controlling chemical reactivity with external electromagnetic fields \cite{Krems:08,Balakrishnan:16,Bohn:17}, the idea that fascinated scientists for decades, and led to the development of new research frontiers at the interface of chemistry and physics, such as mode-selective chemistry \cite{Zare:98,Guo:16}, quantum coherent control \cite{Shapiro:12}, and attochemistry \cite{Nisoli:17}. 
In particular, the production and trapping of ground-state molecular radicals NaLi(a$^3\Sigma^+$),  Li$_2$(a$^3\Sigma^+$), Rb$_2$(a$^3\Sigma^+$), SrF($^2\Sigma^+$), CaF($^2\Sigma^+$), YO($^2\Sigma^+$),  YbF($^2\Sigma^+$) \cite{Rvachov:17, Barry:14,McCarron:18,Anderegg:18,Kozyryev:17,Collopy:18,Lim:18} and studies of their collisional properties  at $\mu$K temperatures \cite{Drews:17,Polovy:20, Cheuk:20,Son:20} suggested the possibility of using the reactants's  electron spin degrees of  freedom to tune ultracold reaction dynamics by magnetic fields.

The prospect of using magnetic fields as a tool to control chemical reactivity is central to ultracold chemistry \cite{Krems:08,Balakrishnan:16} and a very important one in chemical kinetics \cite{Steiner:89} and biological magnetoreception \cite{Hore:16}, where radical pair reactions in cryptochrome proteins are thought to play a key role in magnetic-field-guided orientation of birds and insects \cite{Hiscock:16,Nohr:17}. 
However, despite the long-standing significance of this question and the recent experimental observations of inelastic collisions in  an ultracold Na{-}NaLi(a$^3\Sigma^+$) mixture  \cite{Son:20},
no theoretical studies have been reported on ultracold reaction dynamics involving ground-state alkali-metal dimers and atoms in the presence of external magnetic fields and hyperfine interactions.
This is because such reactions occur through the formation of a deeply bound reaction complex \cite{Althorpe:03,Croft:17,Kendrick:20}, whose numerous strongly coupled bound and resonance states defy rigorous quantum scattering calculations    \cite{Croft:17,Morita:19b,Kendrick:20}.



Here, we explore the dynamics of the ultracold chemical reaction  Na~+~NaLi$(a^3\Sigma^+)$ $\to$ Na$_2(^1\Sigma_g^+)$~+~Li in the presence of magnetic fields and hyperfine interactions using the extended coupled-channel statistical (CCS) model \cite{Tscherbul:20} parametrized by {\it ab initio} calculations. The model assumes the existence of a long-lived reaction complex at short range, whose  properties can be modeled statistically ({\it i.e.} using classical probabilities) \cite{Rackham:03,Alexander:04,Croft:14}. Statistical (or universal) models  \cite{Clary:90,Rackham:03,Alexander:04,Quemener:10,Auzinsh:11,Kotochigova:10,Idziaszek:10,Gao:10,Quemener:12,Buchachenko:12,Gonzalez-Martinez:14,Croft:14,Tscherbul:15b,Frye:15,Yang:20} have been successfully applied to calculate the rate of ultracold chemical reactions of alkali-metal dimers \cite{Quemener:10,Idziaszek:10,Gao:10,Kotochigova:10,Gonzalez-Martinez:14} and the density of states of the (KRb)$_2$ reaction complex \cite{Liu:20}. However,  the previous calculations have been limited to the case of zero magnetic field and did not account for electron spins, hyperfine interactions, and non-adiabatic effects, all of which we will consider in the present work.
 
Our calculations show that the fully spin-polarized spin states of NaLi and Na are $\sim$10-100 times less chemically reactive than unpolarized spin states, demonstrating extensive quantum spin state control of chemical reactions of triplet-state alkali-metal dimers with alkali-metal atoms. We also find that the magnetic field dependence of the reaction rate displays several magnetic Feshbach resonances (MFRs), providing the first theoretical prediction of MFRs in an ultracold chemical reaction. 
MFRs in non-reactive scattering of NaK with K were observed experimentally and thoroughly 
analysed in Refs.~\cite{Rui:17,Yang:19,Wang:21}.
 Our findings open up several new avenues of research in ultracold molecular physics and chemistry. The reactive MFRs will enable experimentalists to efficiently suppress unwanted chemical reactivity in trapped atom-molecule mixtures, enabling, e.g., efficient sympathetic cooling \cite{Son:20,Lara:06,Tscherbul:11,Morita:17,Morita:18,Frye:16,Jurgilas:21}. 
 They could also be used to assemble chemically reactive atom-molecule trimers via magnetoassociation, to engineer entangled many-body states in trapped atom-molecule mixtures, and  to probe and control the quantum dynamics of chaotic scattering and reaction complex formation \cite{Croft:17}.



\begin{figure}
\includegraphics[scale=0.47,trim=40 50 0 65]{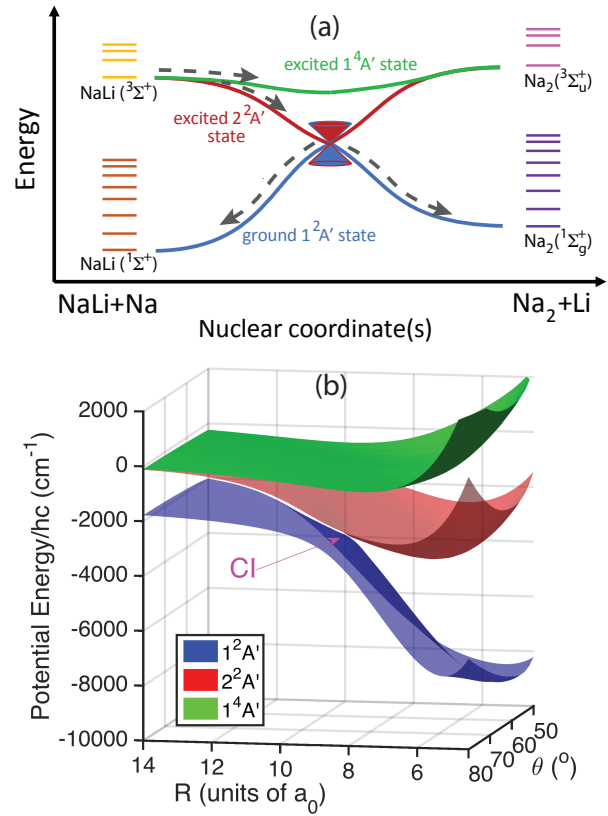}
\caption{(a) Schematic of the Na~+~NaLi(a$^3\Sigma^+$) reactive scattering through a CI between the  $^2\!A^\prime$ PESs leading to either ground state  NaLi(X$^1\Sigma^+$) or Na$_2$(X$^1\Sigma^+_g$) molecules. The CI is indicated by the red/blue cone. (b) {\it Ab initio} adiabatic PESs for Na-NaLi as functions of the Na-to-NaLi  separation $R$ and of the bending angle $\theta$ with $r= 9.1 a_0$, close to the equilibrium distance of the NaLi(a$^3\Sigma^+$) potential. The blue ($1^2A^{\prime}$) and red (2$^2A^{\prime}$) PESs have a CI, where two PESs of the same electronic symmetry touch. The green surface is the spin-polarized, nonreactive PES of the $1^4A'$ symmetry.}
\label{fig:NaNaLi_PESs}
\end{figure}

{\it Theory: Ab initio calculations and extended CCS model.} 
To describe  ultracold  reactive collisions between Na atoms and NaLi molecules in the metastable a$^3\Sigma^+$ electronic state, we performed {\it ab initio} calculations of  the electronic potential energy surfaces (PESs) of the long-lived intermediate Na$_2$Li reaction complex. The complex is characterized by  two $^2A^{\prime}$ and one $^4A^{\prime}$ trimer electronic states. The potential landscape of these barrierless PESs  is shown in Fig.~\ref{fig:NaNaLi_PESs}. The PESs are expressed in the Jacobi coordinates $\mathbf{R}$---the atom-molecule separation vector and $\mathbf{r}$---the vector joining the nuclei of the diatomic molecule.
 For our purposes it is sufficient to determine the PESs, which are only functions of 
$R$ and $\theta$ (the angle between $\mathbf{R}$ and $\mathbf{r}$) in the two-dimensional plane with the internuclear distance of NaLi fixed at its  equilibrium value ($r=r_e$) \cite{Tscherbul:20}.
Our {\it ab initio} calculations of the two-state $^2A'$ PESs  reveal a conical intersection (CI) between the two doublet states which is located  at $R\simeq 8.5 a_0$ and $\theta=70^o$. The relevant multi-dimensional PESs have been determined using the internally-contracted multi-reference configuration interaction (MRCI) method \cite{werner:88} with single and double excitations and Davidson correction~\cite{Langhoff1974a} as  further described in the Supplemental Material \cite{SM}.

From the energetics of the relevant molecular states in the entrance and exit reaction channels we determine that the production of the Na$_2$(a$^3\Sigma^+_u$) molecule in the  Na($^2$S)~+~NaLi(a$^3\Sigma^+$) reaction is endothermic by 
41.7 cm$^{-1}$ including the zero-point vibrational  energy correction.
This suggests that  the vibrational excitation of the reactant NaLi(a$^3\Sigma^+$) molecule to the $v\ge 2$ vibrational states will allow for production of triplet-state Na$_2$ products. However, the CI allows for an efficient transfer of the reactant NaLi(a$^3\Sigma^+$) molecules into either NaLi (X$^1\Sigma^+$) or 
Na$_2$(X$^1\Sigma^+_g$) states of the ground electronic configuration. A schematic depiction of reactive scattering between Na atoms and NaLi(a$^3\Sigma^+$)  molecules through a CI is demonstrated in Fig.~\ref{fig:NaNaLi_PESs}(a).  The reactants start out on the asymptotically degenerate  $2^2A'$ and $1^4A'$ excited PESs. The reaction flux on the  $2^2A'$ PESs can reach the CI and make a transition to the ground $1^2A'$ PES leading to ground-state reaction products. Only the $2^2A'$ and $1^4A'$  PESs are included in our CCS calculations, which is justified by the fact that the CI is located deeply inside the reaction complex region not explicitly included in the calculations \cite{SM}.

The extended CCS model  of  barrierless chemical reactions  \cite{Tscherbul:20}
 assumes the existence of a long-lived reaction complex, whose formation from the reactants or decay to products can be treated as independent events \cite{Rackham:03,Alexander:04}. The state-to-state reaction probability between the reactant and product states $r$ and $p$ 
$P_{r\to p} (E)=\frac{p_p(E)  p_r(E)}{\sum_c p_c(E)},$
where $p_r(E)$ and $p_p(E)$ are the energy-dependent capture probabilities of the reactants and products into the reaction complex obtained by solving the time-independent Schr{\"o}dinger equation in the entrance reaction channel subject to a short-range  capture boundary condition for the reactive $2^2A'$ PES and a regular boundary condition for the nonreactive $^4A'$ PES \cite{Tscherbul:20,SM}.

\begin{figure}
\includegraphics[width=0.9\linewidth, trim = 50 10 -10 20]{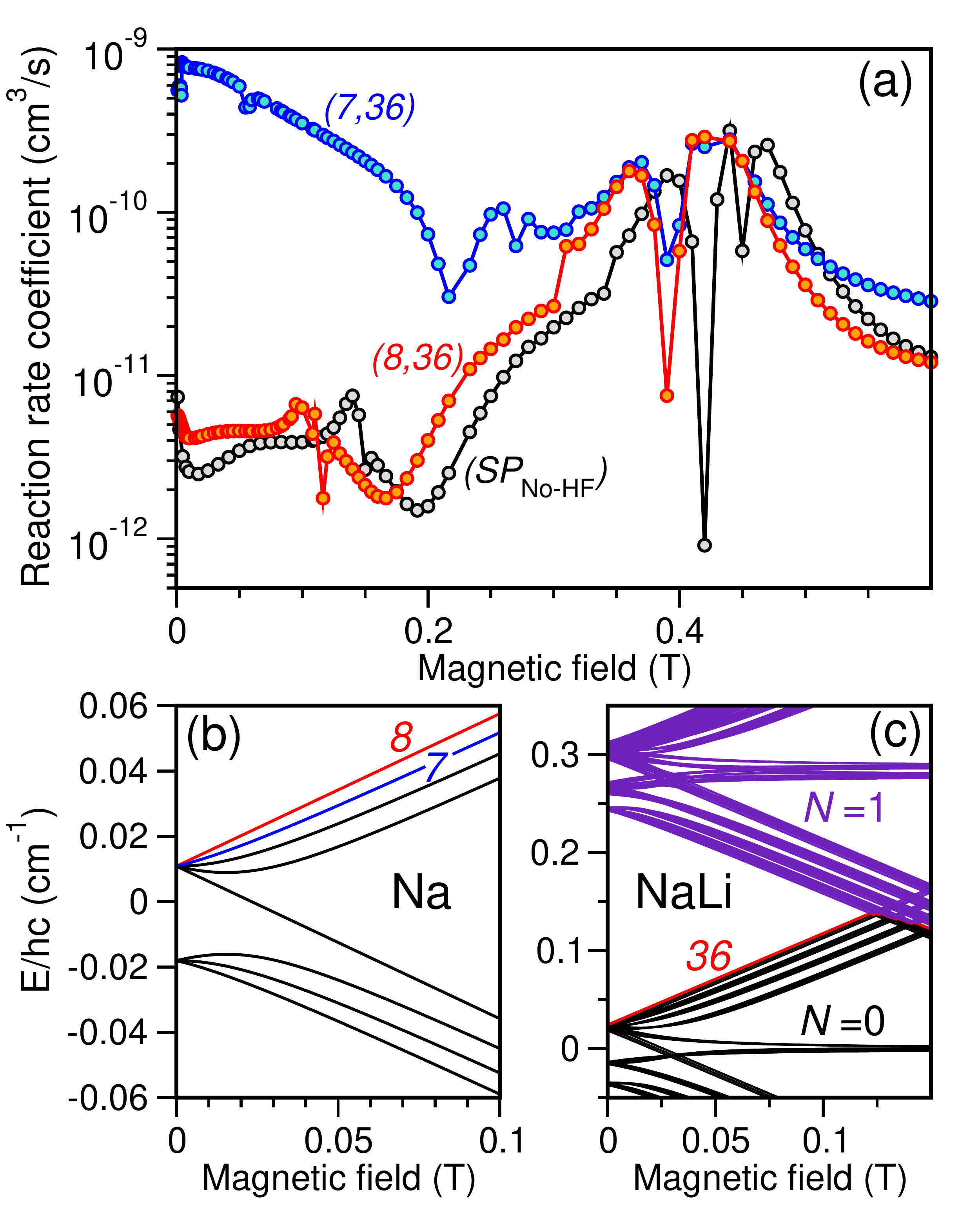}
\caption{(a) Magnetic-field dependence of the reaction rate coefficient for the fully spin-polarized
Na(8)~+~NaLi(36) [red curve with label (8,36)] and non-fully spin polarized Na(7)~+~NaLi(36) collisions [blue
curve].  Also shown are results [black curve labeled by $SP_{\rm No-HF}$] for calculations where the hyperfine
contact interactions are turned off.  Here, the initial state is the spin-polarized
Na($|S_B,M_{S_B}=1/2,1/2\rangle$) + NaLi($|S_A,M_{S_A}=1,1\rangle$) state.  The collision energy is $E/k = 2$ $\mu$K for all data. Here, $k$ is the Boltzmann constant.  Panel (b) shows the hyperfine and Zeeman energy levels of
the ground-state Na atom.  Panel (c) shows the rotational, hyperfine, and Zeeman energy levels of the $N=0$
and 1 rotational states of NaLi(a$^3\Sigma^+,v = 0$). In panels (b) and (c) relevant hyperfine
states (blue and red colored curves) are indexed as $1,2,3,\ldots$ in the order of increasing energy. }
\label{fig:NaLi_Na_levels}
\end{figure}

{\it Ultracold reaction dynamics in a magnetic field.}
We begin by describing the hyperfine energy  level structure of the reactants
in a magnetic field.
 Figures~\ref{fig:NaLi_Na_levels}(b) and (c) show the  Zeeman levels of Na and NaLi($a^3\Sigma^+,v=0,N=0$) obtained by diagonalization of the atomic and molecular Hamiltonians \cite{SM}.
 There are a total of 36 molecular energy levels in the $N=0$ manifold of NaLi(a$^3\Sigma^+$), which can be classified in the weak-field limit by the values of the total angular momentum of the molecule $F$ and its projection on the field axis $M_F$ \cite{Brown:03,Tscherbul:18b}. The calculated zero-field hyperfine splittings are in good agreement with the measured values  \cite{Rvachov:17,SM}.

To explore the influence of reactant spin polarization on chemical reactivity, we consider reactive collisions of NaLi molecules in the highest-energy level $|36\rangle$ of the $N=0$ manifold with Na atoms in the hyperfine states $|7\rangle$ and $|8\rangle$ [see Figs.~\ref{fig:NaLi_Na_levels}(b) and (c)].
Note that state $|36\rangle$ is a triply spin-polarized state of NaLi, where all of the spins in the molecule are aligned along the magnetic field. 
Similarly, state $|8\rangle$ of Na is doubly spin-polarized ($|F=2,m_F=2\rangle$), in contrast to state $|7\rangle$.
    In the absence of the hyperfine structure, the Zeeman states of NaLi and Na shown in Fig.~\ref{fig:NaLi_Na_levels} reduce to 3 molecular states $|S_A M_{S_A}\rangle$ ($M_{S_A}=0,\pm 1$), and 2 atomic states $|S_B M_{S_B}\rangle$ ($M_{S_B}=\pm 1/2$). {The fully spin-polarized initial states of Na and NaLi are labeled as $|2\rangle$ and $|3\rangle$.}

In Fig.~\ref{fig:NaLi_Na_levels}(a) we plot the magnetic field dependence of the reaction rates for the (8,36) and (7,36) initial states of Na~+~NaLi(a$^3\Sigma$) at $T=2$~$\mu$K. The rates are nearly temperature independent, as expected for a two-body inelastic process near an $s$-wave threshold \cite{Balakrishnan:98}.  

More significantly, we observe that the chemical reactivity of fully spin-polarized reactants Na(8)~+~NaLi(36) is suppressed by a factor of $\simeq$10-100  compared to that of non-fully spin-polarized reactants  Na(7)~+~NaLi(36). 
 Remarkably, flipping the  electron spin of one of the reactants leads to a dramatic change in chemical reactivity. While the strong dependence on the initial spin state has been observed previously for Penning ionization in cold atom-atom collisions \cite{Flores:16}, the atom-molecule reaction studied here is essentially different due to the large number of participating rovibrational states coupled by strongly anisotropic atom-molecule interactions. 
 

The rate of the  Na(7)~+~ NaLi(36) reaction displays the opposite trend, beginning to decrease at $B\ge  0.05$~T. This trend is similar to that observed in \cite{Tscherbul:20} and can be explained by referring to Eq. (\ref{Na_state7}): the weight $ c_2(B)$ of the ``reactive'' electron spin state $|\frac{1}{2},-\frac{1}{2}\rangle$ in the hyperfine state  $|7\rangle$ of Na 
\begin{equation}\label{Na_state7}
|7\rangle = c_1(B) | {\textstyle \frac{1}{2} \frac{1}{2}} \rangle |{\textstyle \frac{3}{2} \frac{1}{2}}\rangle + c_2(B) |{\textstyle \frac{1}{2},-\frac{1}{2} } \rangle |{\textstyle \frac{3}{2} \frac{3}{2}}\rangle
\end{equation}
decreases with increasing magnetic field, as the state tends to the unreactive spin-polarized state $|\frac{1}{2}\frac{1}{2}\rangle |\frac{3}{2}\frac{1}{2}\rangle$ in the large-field limit (where $|\frac{1}{2}\frac{1}{2}\rangle |\frac{3}{2}\frac{1}{2}\rangle$  denotes  the Zeeman state with $S_B = M_{S_B}=\frac{1}{2}$,  $I_B =\frac{3}{2}$, and $M_{I_B}=\frac{1}{2}$).
The hyperfine state $|7\rangle$ of Na becomes less and less reactive towards NaLi with increasing field because the reactive weight  $c_2(B)\simeq B^{-1}$  \cite{Tscherbul:20}.  We note that the spin-polarized reaction rates calculated with and without the hyperfine structure of Na and NaLi taken into account [see Fig.~\ref{fig:NaLi_Na_levels}(a)] are similar in magnitude and  field dependence. The fully spin-stretched hyperfine states $|36\rangle$ of NaLi and $|8\rangle$ of Na are direct products of the electron and nuclear spin states, so the nuclear spin degree of freedom only causes a slight shift in threshold energies, but otherwise plays the role of a spectator.


The suppression of chemical reactivity of spin-polarized molecules is due to a general  mechanism  \cite{Moore:73,Tscherbul:06,Abrahamsson:07} based on approximate conservation of the total spin of the reaction complex. 
Specifically, if the electron spins of the reactants are completely polarized, the reaction complex is initialized in the nonreactive state of total spin $S=3/2$ described by the $^4A'$ PES (see Fig.~\ref{fig:NaNaLi_PESs}).
Thus, in the absence of $S$-nonconserving interactions, such as the intramolecular spin-spin or intermolecular magnetic dipole interactions, the value of $S$ must be the same for the reactants and products (the Wigner spin rule \cite{Moore:73}). The energetically allowed  products of the Na~+~NaLi reaction---molecular Na$_2$($^1\Sigma^+_g$) and atomic Li($^{2}$S$_{1/2}$)---correspond to $S=1/2$. 
 As a result, the spin-polarized chemical reaction Na~+~NaLi($a^3\Sigma^+$) $\to$ Na$_2$($^1\Sigma^+_g$)~+~Li requires  spin-changing intersystem crossing transition $S=3/2\to 1/2$ \cite{Marian:12,Lykhin:16,Lykhin:20,Penfold:18,Li:19} in order to proceed. We verified that omitting the  spin-spin and magnetic dipolar interactions from  CCS calculations leads to a complete suppression of the reaction Na(8)~+~NaLi(36) $\to$ Na$_2$~+~Li, while having little effect on the reactivity of the initial state (7,36).

\begin{figure}[t]
\begin{center}
\includegraphics[width=0.8\linewidth, trim = 20 40 0 -10]{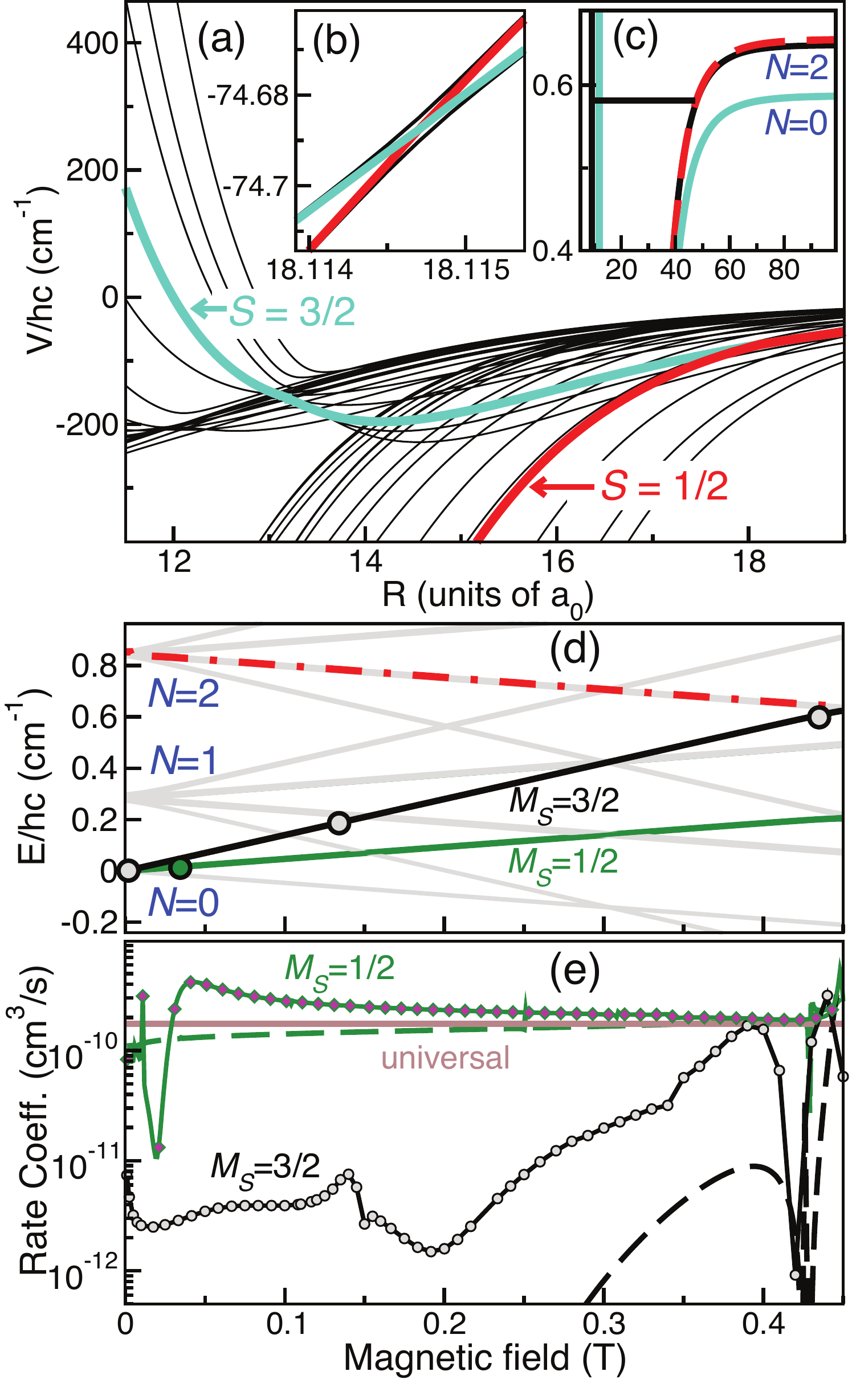}
\end{center}
\caption{(a) Adiabatic potentials (thin black curves) of the Na-NaLi reaction complex in the absence of the hyperfine
contact interactions at $B = 0.01$ T and $M = 3/2$. 
Diabatized potentials for $S=1/2$ and $3/2$ are shown as cyan and red curves, respectively.
(b) A blowup of the avoided crossing near $18.1 a_0$; (c) Open-channel (cyan curve) and closed-channel (black and red dashed curves) diabatic potentials near the $N=0$ and 2 NaLi rotational thresholds at $B = 0.42$~T. The closed-channel bound state is shown by the horizontal bar.  (d) Threshold energies (grey curves) of Na+NaLi as functions of magnetic field. Our incident thresholds labeled $M_S=1/2$ and $M_S=3/2$ are colored as green and black curves, respectively ($M_S=M_{S_A}+M_{S_B}$). The $|N_A=2,M_{N_A}=-1,M_{S_A}=0\rangle|M_{S_B}=-\frac{1}{2}\rangle$  closed-channel threshold is the dashed red curve; (e)  Na + NaLi reaction rate coefficients for the spin-polarized $M_S=3/2$ (circles) and unpolarized $M_S=1/2$ (diamonds) initial states as functions of magnetic field.  Solid and dashed lines correspond to calculations including and excluding the anisotropic part of the Na-NaLi PESs. The universal limit is indicated by the brown horizontal line.  Locations of MFRs in this panel and thus of zero-energy closed-channel bound states, are shown as colored circles in panel (d).} 
\label{fig:avoided_crossings}
\end{figure}

To gain further insight into the mechanism of the spin-polarized chemical reaction Na~+~NaLi$(a^3\Sigma^+)$ we plot in Fig.~\ref{fig:avoided_crossings}(a) the adiabatic eigenvalues $\epsilon_i(R)$ of the atom-molecule Hamiltonian  \cite{Clary:90,Auzinsh:11,Aquilanti:82,Aquilanti:99,Nobusada:99}. 
 Consider, e.g., the $S=3/2$ diabatic potential
    obtained by following the corresponding adiabatic curves through a series of avoided crossings shown in  Fig.~\ref{fig:avoided_crossings}. The potential  is repulsive at short  range with a well depth of $\simeq 200$ cm$^{-1}$, and correlates with the fully spin-polarized  initial state of Na(2)-NaLi(3). 
    The repulsive state experiences several crossings with the $S=1/2$ diabatic states, which are attractive at short range and correlate asymptotically with unpolarized rotationally excited states of NaLi. 
     The crossings are induced by $S$-nonconserving interactions, predominantly by the intramolecular spin-spin interaction of NaLi$(a^3\Sigma)$, which cause the chemical reaction.
 We  note  that  a simple two-channel model involving the pair of diabatic states near the largest avoided crossing shown in Fig.~\ref{fig:avoided_crossings}(b)
underestimates the reaction rate by several orders of magnitude (as does Landau-Zener theory), suggesting the importance of multichannel effects.


The resonance variation of the spin-polarized reaction rate near $B=0.4$~T shown in Fig.~\ref{fig:NaLi_Na_levels}(a)

 is caused by MFRs, which occur due to the coupling of the incident spin-polarized  channel $|N_A=0,M_{S_A}=1\rangle$  with closed-channel bound states $|N_A'=2,M_{S_A}'\rangle$ ($M_{S_A}'\ne M_{S_A}$) mediated by {\it anisotropic interactions}, which include the intramolecular spin-spin interaction of NaLi(a$^3\Sigma^+)$ \cite{Krems:04} and the anisotropic part of the Na-NaLi interaction.
 The near-threshold bound state responsible for the MFR at 0.42~T is supported by the adiabatic potential that correlates to the  $|N_A'=2,M_{S_A}'=0 \rangle|M_{S_B}'=-\frac{1}{2}\rangle$  closed-channel threshold, as shown in Fig.~\ref{fig:avoided_crossings}(c).

 Figure~\ref{fig:avoided_crossings}(e) illustrates that MFRs can also occur in the spin-unpolarized incident channel (1,3).  The low-field resonance is mainly due to the atom-molecule interaction anisotropy, which couples the $N=0$ incident channel with $N>0$  closed channels. Indeed, as shown in Fig.~\ref{fig:avoided_crossings}(e) the MFR disappears when the anisotropic part of  the  Na-NaLi interaction is omitted.

Our calculated Na-NaLi reaction rates deviate substantially from the universal value $k_{0}^u=1.84\times 10^{-10}$ cm$^3$/s \cite{Julienne:11,Li:19b}  calculated using the accurate {\it ab initio} Na-NaLi(a$^3\Sigma^+$) long-range dispersion coefficient $C_6 = 4026$~a.u. \cite{SM}. This indicates a substantial degree of non-universality due to the inherently multichannel nature of the reaction dynamics caused by anisotropic interactions (see above).
 As shown in Fig.~\ref{fig:avoided_crossings}(a) a large fraction of adiabatic channels, through which the reaction occurs, is repulsive at short range,  leading to a significant reflection of the incident flux even for unpolarized initial reactant states. This reflection manifests in the appearance of MFRs and other non-universal effects \cite{Julienne:11}.
 Test calculations show that in the absence of anisotropic interactions, the unpolarized reaction rate remains close to the universal limit over the entire range of magnetic fields [see Fig.~\ref{fig:avoided_crossings}(e)].

\color{black}

In summary, we have presented a theoretical study of the ultracold chemical reaction of Na atoms with triplet  NaLi(a$^3\Sigma^+$) molecules in their ground rovibrational states in the presence of external magnetic fields and hyperfine interactions. This reaction is representative of a wide class of ultracold chemical reactions of triplet alkali-dimer molecules  currently studied by several experimental groups \cite{Son:20,Drews:17,Polovy:20}.
Our calculations reveal a substantial  degree of quantum state selectivity in the dependence of the reaction rate on the initial states of the reactants (fully spin-polarized vs. unpolarized). 
 Our results also suggest  that it is possible to control ultracold chemical reactions of alkali-metal dimers with alkali-metal atoms via magnetic Feshbach resonances. 

The generality of the spin-based control mechanisms explored here implies their potential utility as a tool to control other, potentially more complex chemical reactions, such as those of heavier bialkali molecules [e.g.,  K~+~KRb(a$^3\Sigma$)] and those  involving $^2\Sigma$ molecules, such as Li~+~CaH($^2\Sigma$) \cite{Tscherbul:20,Tscherbul:11}, Li~+~SrOH($^2\Sigma$)  \cite{Morita:17}, and  Li~+~CaF($^2\Sigma$) \cite{Frye:16}. 
We thus expect our results to be tested in near-future experiments with ultracold atom-molecule mixtures. 

{We are grateful to Wolfgang Ketterle, Hyungmok Son, Alan Jamison, and Sergey Varganov for stimulating discussions.}
Work at the University of Nevada, Reno was supported by the NSF Grant  No. PHY-1912668. Work at Temple University is supported by the Army Research Office Grant No. W911NF-17-1-0563 and the NSF Grant No. PHY-1908634.

\end{document}